\documentclass[a4paper,11pt]{book}
\usepackage{graphicx}
\usepackage{color}
\usepackage{amsmath,amsfonts}

\newcommand{\ie}{{\it i.e.}}
\newcommand{\dd}{{\rm d}}
\newcommand{\ii}{{\rm i}}
\newcommand{\ee}{{\rm e}}
\newcommand{\re}{{\rm Re}}
\newcommand{\dx}{\dd^3 x}
\newcommand{\dk}{{\dd^{3}k}}
\newcommand{\om}{\omega}
\newcommand{\bx}{{\bf x}}
\newcommand{\bk}{{\bf k}}

\newcommand{\pdt}{\partial_t}
\newcommand{\tra}{{\rm T}}

\newcommand{\hH}{\hat H}
\newcommand{\hN}{\hat N}

\newcommand{\nx}{\nabla_{\bf x}}

\newcommand{\hP}{\hat\Psi}
\newcommand{\hPd}{\hat\Psi^\dagger}
\newcommand{\Pn}{\Psi_0}
\newcommand{\rn}{\rho_0}
\newcommand{\rnx}{\rho_0(x)}
\newcommand{\hp}{\hat\phi}
\newcommand{\hpd}{\hat\phi^\dagger}
\newcommand{\hpds}{\hat\phi^{\dagger 2}}

\newcommand{\hW}{\hat W}

\newcommand{\uk}{u_\bk}
\newcommand{\vk}{v_\bk}
\newcommand{\vks}{v_\bk^*}

\newcommand{\ak}{\hat a_\bk}
\newcommand{\akd}{\hat a_\bk^{\dagger}}
\newcommand{\akdp}{\hat a_{\bk'}^{\dagger}}

\newcommand{\hbm}{\frac{\hbar^2}{2m}}

\newcommand{\vev}[1]{{\langle 0 | #1 | 0 \rangle}}
\newcommand{\sev}[1]{{\langle \zeta | #1 | \zeta \rangle}}
\newcommand{\comm}[2]{[#1,#2]}
\newcommand{\spin}[2]{\begin{pmatrix} {#1} \cr {#2}\end{pmatrix}}
\newcommand{\scal}[2]{{\langle #1 | #2 \rangle}}

\title{Proceedings of the II Amazonian Symposium on Physics}

\begin{document}

\maketitle

\chapter*{The analogue cosmological constant in Bose--Einstein condensates: a lesson for quantum gravity}

\author{Stefano Finazzi\footnote{finazzi@science.unitn.it}\\
\textit{INO-CNR BEC Center and Dipartimento di Fisica, Universit\`a di Trento, via Sommarive 14, 38123 Povo--Trento, Italy}\\
Stefano Liberati\footnote{liberati@sissa.it}\\
\textit{SISSA, Via Bonomea 265, 34136, Trieste, Italy and INFN, Sezione di Trieste}\\
\and
Lorenzo Sindoni\footnote{lorenzo.sindoni@aei.mpg.de}\\
\textit{Albert Einstein Institute, Am M\"uhlenberg 1, 14476 Golm, Germany}\\
}

\textbf{Abstract}:
For almost a century, the cosmological constant has been a mysterious object, in relation to both its origin and its very small value.
By using a Bose--Einstein condensate analogue model for gravitational dynamics, we address here the cosmological constant issue from an analogue gravity standpoint.
Starting from the fundamental equations describing a system of condensed bosons, we highlight the presence of a vacuum source term for the analogue gravitational field, playing the role of a cosmological constant. In this simple system it is possible to compute from scratch the value of this constant, to compare it with other characteristic energy scales and hence address the problem of its magnitude within this framework, suggesting a different path for the solution of this longstanding puzzle.
We find that, even though this constant term is related with quantum vacuum effects, it is not immediately related to the ground state energy of the condensate. On the gravity side this result suggests that the interpretation and computation of the cosmological term as a form of renormalized vacuum energy might be misleading, its origin being related to the mechanism that instead produces spacetime from its pregeometric progenitor, shedding a different light on the subject and at the same time suggesting a potentially relevant role of analogue models in the understanding of quantum gravity.

\section{Introduction}

The cosmological constant~\cite{carroll} has been one of the most mysterious and fascinating objects for both cosmologist and theoretical physicists since its introduction almost one century ago~\cite{firstcosm}. Once called by Einstein his greatest blunder, it seems nowadays the driving force behind the current accelerated expansion of the universe. At the theoretical level, the explanation of its origin is considered one of the most fundamental issues for our comprehension of general relativity (GR) and quantum field theory (QFT).

Since this constant appears in Einstein's equations as a source term present even in the absence of matter and with all the symmetries of the vacuum (\ie~a stress-energy tensor of the form $T_{\mu\nu}^\Lambda\propto g_{\mu\nu}$), it is usually interpreted as a vacuum energy, essentially related to the zero point fluctuations of quantum fields. This reasonable point of view has originated what is often called the ``worst prediction of physics''. Indeed, its theoretical value, which is na\"ively obtained by integrating the zero-point energies of modes of quantum fields below Planck energy (but can be computed also by more sophisticated renormalization arguments), is about 120 orders of magnitude larger than the measured value. {We can summarize the situation by saying that, given the absence of custodial symmetries protecting the cosmological term from large renormalization effects, the only option we have to explain observations is fine tuning~\cite{finetuning, rovelli}.}

Despite the large number of attempts (most notably supersymmetry~\cite{susy}, which, however, must be broken at low energy) to improve this estimate, this problem is still waiting for a complete and satisfactory solution. 

This huge discrepancy is plausibly due to the use of effective field theory (EFT) calculations for a quantity which can be computed only within a full quantum gravity (QG) theory (see, however, \cite{Paddy} for a proposal in the semiclassical gravity limit). Unfortunately, to date, we do not have any conclusive theory at our disposal, allowing a complete calculation of the gravitational effective action from basic principles, and solving the naturalness problem associated to the cosmological constant. 

Therefore, in order to give support to this idea, some other arguments have to be presented. Analogue models for gravity~\cite{livrev}, in this respect, give us the possibility to study the emergence of given EFT with certain geometrical content from microscopic constitutents (typically interacting atoms, fluids, etc.), keeping the process under control at every stage of the transition, and hence allowing us to show at which point of the calculation the EFT intuition of the cosmological constant as a vacuum energy term fails. 

Already in~\cite{volovik1,volovik2,volovikbook} it was shown that a na\"ive computation of the ground state energy using the EFT (the analogue that one would do to compute the cosmological constant) would produce a wrong result. The unique way to compute the correct value seems to start from the full microscopic theory, working out the macroscopic quantities from it.

Given the deep difference in the structure of the equations of fluid dynamics and those of GR (and other gravitational theories) it is not possible to have an accurate analogy at the dynamical level: indeed, this is forbidden by the absence of diffeomorphism invariance. However, in~\cite{gravdynam} it has been shown for the first time that the evolution of the acoustic metric in a Bose--Einstein condensate (BEC) is described by a Poisson equation for a nonrelativistic gravitational field, thus realizing a (partial) dynamical analogy with Newtonian gravity. Noticeably, this equation is endowed with a source term which is present even in the absence of real phonons and can be naturally identified as a cosmological constant.

In this chapter we will consider such analogue model for gravity and directly show that the cosmological constant term cannot be computed through the standard EFT approach~\cite{cosmobec}, confirming the conjecture of~\cite{volovik1,volovik2}. Somehow unexpectedly, we find that also the total ground state energy of the condensate does not give the correct result: indeed, the cosmological constant is comparable with that fraction of the ground state energy corresponding to the quantum depletion of the condensate, \ie\/ to the fraction of atoms inevitably occupying excited states of the single particle Hamiltonian.

This result is twofold. First, it gives an explicit calculation showing qualitatively and quantitatively where the EFT intuition might fail to grasp the nature of the vacuum term. Second, it shows that the subject of analogue models is not just a mere curiosity for condensed matter physics, but it is also of great interest for research in quantum gravity, for their ability in providing guidance and patterns to be used to address what is the most urgent problem there, that is the recovery of the continuum semiclassical limit.

\sectionmark{The problem in EFT}			%
\section{The cosmological constant problem in EFT}	%
\label{sec:effective field theory}			%
\sectionmark{The problem in EFT}			%

The cosmological constant $\Lambda$ enters Einstein's equation
\begin{equation}
 R_{\mu\nu}-\frac{1}{2}g_{\mu\nu}R+\Lambda\,g_{\mu\nu}=\frac{8\pi G}{c^4}T_{\mu\nu}
\end{equation}
as a term multiplying the metric tensor $g_{\mu\nu}$.
This means that we can interpret $\Lambda$ as a vacuum energy density
\begin{equation}
 {\cal E}_\Lambda=\frac{c^4\Lambda}{8\pi G}
\end{equation}
and hence see the cosmological term as a vacuum stress energy term
\begin{equation}
 T_{\mu\nu}^{\Lambda}=-{\cal E}\,g_{\mu\nu}.
\end{equation}
%

The other way around, this argument implies that every vacuum energy is in the form of a cosmological constant term and must be thus identified with $\Lambda$. 

The problem arises when computing the renormalization of this term using quantum field theory techniques. Summing up all the zero-point energies of all normal modes of some quantum field of mass $m$ up to a cutoff energy $\mu$, one obtains~\cite{weinbergcc}
\begin{equation}
 {\cal E}=\int_0^{\mu/\hbar c}\frac{4\pi\,k^2\dd k}{(2\pi)^3}\,\frac{1}{2}\hbar c\sqrt{k^2+\frac{m^2 c^2}{\hbar^2}}
 \approx\frac{\mu^4}{16\pi^2(\hbar c)^3}.
\end{equation}
If we compute this term with a Planck-scale cutoff, we obtain~\cite{carroll} an estimate for the vacuum energy of the order of
\begin{equation}
 {\cal E}_{\rm P}\approx 10^{110}\,\mbox{erg}/\mbox{cm}^3,
\end{equation}
while if we lower the cutoff to a much smaller energy scale, like the QCD scale, we expect 
\begin{equation}
 {\cal E}_{\rm QCD}\approx 10^{36}\,\mbox{erg}/\mbox{cm}^3.
\end{equation}
The cosmological observations give
\begin{equation}
 {\cal E}_{\rm obs}\approx 10^{-10}\,\mbox{erg}/\mbox{cm}^3, 
\end{equation}
a value hardly explained in terms of obvious particle physics scales.

In a proper semiclassical renormalization procedure of the whole theory (gravity plus quantum fields) a suitable bare cosmological constant $\Lambda_{\rm b}$ has to be introduced to tune the renormalized $\Lambda$ to match its observed value~\cite{birreldavies}. However, since the order of magnitudes of $\Lambda_{\rm  EFT}$ and the observed one are so different, such a procedure would require an extremely fine tuning of $\Lambda_{\rm b}$.

Furthermore, another obscure issue should be clarified. In quantum field theory in Minkowski spacetime, the zero-point energy has no particular meaning, since energy does not gravitate. Actually, it can be safely removed by normal ordering. On the contrary, when gravity is turned on, this is no longer possible, since energy directly enters Einstein's equations: besides the very difficulty of defining a normal ordering prescription in curved dynamical spacetimes, different zero-point energies yield different stress-energy tensors $T_{\mu\nu}$ and, therefore, different solutions of Einstein's equations.

What is then the correct vacuum energy that must be put into Einstein's equations? 

From such considerations, it seems that there is no way out to this problem. The only possibility to compute $\Lambda$ would be to directly know the underlying QG theory from which both GR and EFT emerge as a low energy limit. Under this hypothesis there is no hope of computing $\Lambda$ from scratch working only in a semiclassical gravity framework. In a certain sense $\Lambda$ is an emergent quantity that parametrizes in one number certain details of the microscopic structure and dynamics of  spacetime and influences the macroscopic equations of GR and EFT, and, as such, cannot be justified avoiding a discussion of the microphysics.

As an example of this emergence mechanism, let us mention the propagation of phonons in BECs. In that case, the equation of motion of phonons can be written using only microscopic quantities (speed of sound $c$, velocity $v$, and density of the fluid $\rho$) plus the healing length
\begin{equation}
 \xi\equiv\frac{\hbar}{\sqrt{2}mc}
\end{equation}
which, through the combination of microscopic quantities ($m$ is the mass of bosons), defines the scale at which the dispersion relation
\begin{equation}
 \om^2=c^2 k^2\left(1+\frac{k^2\xi^2}{2}\right)
\end{equation}
is no more in the relativistic regime $\om^2=c^2k^2$. If an observer could only make measures of phonons, he/she would measure the healing length, but he/she would not be able to compute it from first principles, just because he/she would not know the whole theory of BECs but only the phonon EFT.

To shed some light on this problem, it is interesting to have a toy model of gravity where one can compute from first principles both the vacuum energy and the cosmological constant. In this way one can compare them and check if they are the same quantity or they are unrelated instead. In order to do this, one needs a model where it is possible to derive not only the dynamics of a field leaving in an effective geometry described by some metric $g_{\mu\nu}$, as usual in analogue models~\cite{livrev}, but also the equation governing the dynamics of the metric itself. That is, analogue Einstein's equations are needed, so that the analogue cosmological constant can be directly read from them. Such a model was studied in~\cite{gravdynam}, using a BEC with $U(1)$-symmetry breaking. The computation of the analogue cosmological constant is performed in~\cite{cosmobec} and reported in Sec.~\ref{sec:cosmobec}, together with a revision of the model of~\cite{gravdynam}.

\section{Volovik's proposal}	%
\label{sec:volovik}		%

The idea of using analogue models to understand the origin of the cosmological constant is not original of~\cite{cosmobec} but was firstly developed by Volovik~\cite{volovik1,volovik2} for a quantum liquid and presented in~\cite{volovikbook} also for a Bose gas. However, his approach is completely different from that of~\cite{cosmobec}. It is worth briefly reviewing it, to make a comparison with ours.

Even if Volovik did not have an analogue model with an equation describing the dynamics of the geometry as in~\cite{gravdynam}, he nevertheless used a nice argument to determine the analogue cosmological constant.
His argument is based on the identification of the proper thermodynamical potential for the particular considered problem. In this case, we are interested in the emergence of an analogue QFT in condensed matter. The many-body system of identical atoms constituting the quantum liquid is described by the grand canonical Hamiltonian
\begin{equation}\label{eq:volovikH}
 {\cal \hH}=\hH-\mu\hat N,
\end{equation}
where $\hH$ is the second-quantized Hamiltonian, $\mu$ is the chemical potential, and $\hat N$ the particle number operator.

The correct vacuum energy density for the QFT emerging in the many-body system (corresponding to the analogue cosmological constant) is therefore the expectation value of Eq.~\eqref{eq:volovikH} on a state $|0\rangle$ with no phonons, in the thermodynamic limit in which both the volume $V$ and the particle number $N$ goes to infinity
\begin{equation}
 {\cal E}_{\rm vac}=\frac{1}{V}\vev{\hH-\mu\hat N}.
\end{equation}
Using the Gibbs--Duhem relation of thermodynamics~\cite{huang}, stating that at thermodynamic equilibrium
\begin{equation}
 E-TS-\mu N=-pV,
\end{equation}
where $E$ and $N$ are the expectation value of $\hH$ and $\hat N$, respectively, $T$ is the temperature of the system, $S$ its entropy, and $p$ its pressure, one obtains
\begin{equation}\label{eq:statevolovik}
 {\cal E}_{\rm vac}=-p,
\end{equation}
because $T=0$ in the phonon ground state $|0\rangle$.

Equation~\eqref{eq:statevolovik} is the key result of this analysis, since it exactly represents the correct equation of state for the cosmological constant. To summarize, if the vacuum energy is the density of the expectation value of the grand canonical Hamiltonian on the zero-temperature state (no excitations), then $p=-{\cal E}_{\rm vac}$ by thermodynamic relations.
The second interesting feature is that, if $\cal H=\langle{\cal\hH}\rangle$ of Eq.~\eqref{eq:volovikH} were really the quantity that gravitates in place of the energy $E=\langle\hH\rangle$, the freedom in the choice of the zero-point energy would not affect the gravitating quantity $\cal H$. If the energy of each atom were shifted by a factor of $\alpha$, the Hamiltonian would be shifted by $\alpha\hat N$. However, also the chemical potential would have to be shifted of $\alpha$, such that the grand canonical Hamiltonian, being the difference of $\hH$ and $\mu\hat N$, would not change under this transformation. In so doing, there would be a definite gravitating quantity, allowing at the same time for the freedom in the choice of the zero-point energy.

Unfortunately, this argument does not use any dynamical equations for the spacetime geometry. This is, in our opinion, the weakness of the above treatment. Indeed, the cosmological constant is the quantity that gravitates in the absence of matter, encoding the microscopic properties of the spacetime structure. It may be a vacuum energy of some field, but this interpretation may be also wrong. As discussed in the previous section, the only way to obtain a reliable result for the cosmological constant would be to derive a dynamical equation for the metric, from which $\Lambda$ might be directly read off.

\section{A lesson from BECs}	%
\label{sec:cosmobec}		%

As we have argued above, the only safe way to compute the cosmological constant would be to know the microscopic structure of the spacetime and to derive Einstein's equations from it. In this section we apply this procedure to a particular analogue system, a BEC with $U(1)$ breaking~\cite{gravdynam}, for which an equation describing the analogue gravitational dynamics exists.

\subsection{Settings}				%
\label{subsec:settings}				%

The model used in~\cite{gravdynam} is a modified BEC system including a soft breaking of the $U(1)$ symmetry associated with the conservation of particle number. This unusual choice is a simple trick to give mass to quasiparticles that are otherwise massless by Goldstone's theorem. In second quantization, such a system is described by a canonical field $\hPd$, satisfying
\begin{equation}\label{eq:comm}
[\hP(t,\bx),\hPd(t,\bx')]=\delta^{(3)}(\bx-\bx'),
\end{equation}
whose dynamics is generated by the grand canonical Hamiltonian
\begin{equation}
{\cal\hH}=\hH-\mu\hat N,\label{eq:H}
\end{equation}
where
\begin{equation}
 \hH = \int\! \dx \left[\frac{\hbar^2}{2m} \nx \hPd \, \nx\hP + V  \hPd\hP 
 + \frac{g}{2}\hPd\hPd\hP\hP-\frac{\lambda}{2}\left(\hP\hP+\hPd\hPd\right)\right]\label{eq:HU1breaking}
\end{equation}
is the Hamiltonian and
\begin{equation}
 \hat N=\int \!\dx\,\hPd\hP
\end{equation}
is the standard particle number operator for $\hP$.
In the Hamiltonian, $g$ is the coupling constant of the two-body interaction, while $\lambda$, having dimensions 
of energy, represent a $U(1)$ breaking term, associated to a violation of the conservation of the number operator.
For further details on this model and on possible physical realizations, see \cite{gravdynam}. See also \cite{nbec} for a generalization to condensates with many components.

We describe the formation of a BEC at low temperature through the complex function $\Pn$ for the condensate and the operator $\hp$ for the perturbations on top of it, defined by 
\begin{equation}\label{eq:defphicc}
 \hP=\Pn(1+\hp).
\end{equation}
The canonical commutation relation is directly obtained from the commutation rules of the boson field $\hP$ [Eq.~\eqref{eq:comm}]:
\begin{equation}
 \comm{\hp(t,\bx)}{\hpd(t,\bx')}=\frac{1}{\rn(\bx)}\delta^{(3)}(\bx-\bx').
\end{equation}
%
Using Eq.~\eqref{eq:defphicc}, it is convenient to expand the grand canonical Hamiltonian $\cal\hH$ of Eq.~\eqref{eq:H} up to second order in $\hp$
\begin{equation}\label{eq:Hexpansion}
{\cal\hH}\approx{\cal H}_0+{\cal\hH}_1+{\cal\hH}_2,
\end{equation}
where
\begin{align}
{\cal H}_0&= 
 		 \int\!\dx \,\left[\Pn^*\left(- \hbm\nx^2+V-\mu+\frac{g}{2}\rn\right)\Pn-\frac{\lambda}{2}\left(\Pn^2+{\Pn^*}^2\right)\right], \label{eq:h0U1breaking}\\
{\cal\hH}_1	
 	&= \int\!\dx \,\left[\Pn^*\hpd\left(- \hbm\nx^2+V-\mu+g\rn\right)\Pn-\lambda{\Pn^*}^2\hpd\right]+{\rm h.c.}, \label{eq:h1U1breaking}\\
{\cal\hH}_2	
 	&=
\int\! \dx\,\rn\left\{\hpd\left[T_\rho-\ii v\hbar\nx-\hbm\frac{\nx^2\Pn}{\Pn}+V-\mu+2g\rn\right]\hp
\right.\nonumber\\ &\qquad\qquad\qquad\qquad\left.
	+ \frac{\rho}{2}\left(\hpds+\hp^2\right)+\frac{\lambda}{2\rn}\left(\Pn^2{\hp}^2+{\Pn^*}^2{\hpds}\right)
\right\}.
\label{eq:h2U1breaking}
\end{align}

For a stationary condensate, $\pdt\Pn=0$ and the grand canonical Hamiltonian~\eqref{eq:h1U1breaking} generates a modified Gross--Pitaevskii equation
\begin{equation}
  \left[-\hbm\nx^2+V-\mu+g\rn-\lambda\frac{\Pn^*}{\Pn}\right]\Pn=0. \label{eq:GPcc}
\end{equation}
Moreover, to compute the analogue cosmological constant, it is enough to consider homogeneous backgrounds. Thus, one can assume that $V=0$ and the condensate is at rest, such that $\Pn$ has a constant phase, that one can put to $0$ ($\Pn^*=\Pn=\sqrt{\rn}$). With these assumptions, Eq.~\eqref{eq:GPcc} simplifies to
\begin{equation}\label{eq:mu}
 \mu=g\rn-\lambda.
\end{equation}
Under the same assumptions, the equation of motion of the quasiparticles, generated by the second order Hamiltonian~\eqref{eq:h2U1breaking}, reads
 \begin{equation}
   \ii\hbar\pdt\hp = \left[-\frac{\hbar^2}{2m} \nx^2 + g\rn+\lambda
\right]
\hp + \left(g\rn-\lambda\right)
\hpd.\label{eq:hp}
\end{equation}
To solve this equation it is convenient to define at first a two-component field~\cite{ulf}
\begin{equation}
 \hW\equiv\spin{\hp}{\hpd}.
\end{equation}
%
Then Eq.~\eqref{eq:hp} can be written in a compact form
\begin{eqnarray}
 & \ii\hbar\pdt\hW = B\hW,\label{eq:spineq}\\
 & B = (T+g\rn+\lambda)\sigma_3 +\ii \left(g\rn-\lambda\right)\sigma_2,
\end{eqnarray}
where $T$ is the kinetic energy operator
\begin{equation}
 T\equiv-\hbm\nx^2
\end{equation}
and $\sigma_i$ are the Pauli matrices
\begin{equation}
 \sigma_1=
  \begin{pmatrix}
  0 & 1 \\
  1 & 0
  \end{pmatrix}
 ,
 \quad
 \sigma_2=
 \begin{pmatrix}
  0 & -\ii\\
  \ii & 0
 \end{pmatrix},
 \quad
 \sigma_3=
 \begin{pmatrix}
  1 & 0\\
  0 & -1
  \end{pmatrix}
 .
\end{equation}
%
%
Since the field $\hW$ is invariant under the conjugation operation defined by
\begin{equation}
 \bar{S}\equiv\sigma_1 S^\star,
\end{equation}
the structure of $\hW$ must be
\begin{equation}
 \hW=\int\dk (W_\bk \ak + \bar W_\bk \akd),
\label{eq:hW}
\end{equation}
where $W_\bk$ is a doublet of $\mathbb{C}$-functions. 

Using $\nx^\tra=-\nx$, $T^\tra=T$ and the properties of 
Pauli matrices, one verifies 
that the scalar product 
%
\begin{equation}\label{eq:scalar}
 \scal{W_1}{W_2}\equiv\int\!\dx\,\rnx \, W_1^{*\tra}(t,x)\sigma_3 W_2(t,x)
\end{equation}
is conserved under time evolution when $W_i$ are solution of~\eqref{eq:spineq}, since
\begin{equation}
 B^{*\tra}\sigma_3=\sigma_3 B. 
\end{equation}

Imposing the following normalization for the modes
\begin{eqnarray}
 \scal{W_\bk}{W_{\bk'}} &=& -\scal{\bar W_\bk}{\bar W_{\bk'}}=\delta^{(3)}(\bk-\bk'),\label{eq:wnorm}\\
 \scal{W_\bk}{\bar W_{\bk'}} &=& 0,
\end{eqnarray}
one gets
\begin{equation}
 \comm{\ak}{\akdp}=\comm{\scal{W_\bk}{\hW}}{-\scal{\bar W_{\bk'}}{\hW}}=\scal{W_\bk}{W_{\bk'}}=\delta^{(3)}(\bk-\bk'),
\end{equation}
which shows that $\ak$ and $\akd$ are in fact destruction and creation operators.
Moreover, all the other scalar products and all the other commutators vanish.

By homogeneity and stationarity, the doublets must have the following form
\begin{equation}
 W_\bk=\frac{\ee^{-\ii\om t+\ii\bk\cdot\bx}}{\sqrt{\rn(2\pi)^3}}\spin{\uk}{\vk},
\end{equation}
where $\sqrt{\rn(2\pi)^3}$ is a convenient normalization factor and $\uk$ and $\vk$ are constant.
The field $\hp$ can now be expanded as
\begin{equation}\label{eq:phiexpansionk}
\hp
=\int\!\frac{\dk}{\sqrt{\rn(2\pi)^3}}\left[\uk\ee^{-\ii\omega t+\ii{\bf k}\cdot {\bf x}}\ak+\vks\ee^{+\ii\omega t-\ii{\bf k}\cdot {\bf x}}\akd\right]
\end{equation}
%
and the Bogoliubov coefficients $\uk$ and $\vk$ obey the standard normalization
\begin{equation}\label{eq:normuv}
 |\uk|^2-|\vk|^2=1,
\end{equation}
which follows by inserting the above expansion of $\hp$ in Eq.~\eqref{eq:wnorm}. In term of these Fourier components, Eq.~\eqref{eq:hp} reads
\begin{equation}\label{eq:uvequations}
\begin{aligned}
 &\left[\hbar\om-\left(\frac{\hbar^2\bk^2}{2m}+g\rn+\lambda\right)\right]\uk=\left(g\rn-\lambda\right)\vk,\\
 &\left[\hbar\om+\left(\frac{\hbar^2\bk^2}{2m}+g\rn+\lambda\right)\right]\vk=-\left(g\rn-\lambda\right)\uk.
 \end{aligned}
\end{equation}
By imposing that this system has nontrivial solutions, one obtains the quartic dispersion relation
\begin{equation}\label{eq:disp}
 \hbar^2\omega^2=4\lambda g\rn+\frac{g\rn+\lambda}{m}\hbar^2k^2+\frac{\hbar^4k^4}{4m^2},
\end{equation}
describing massive phonons with ultraviolet corrections, mass $\cal M$
\begin{equation}
 {\cal M}=\frac{2\sqrt{\lambda g\rn}}{g\rn+\lambda}m,\label{eq:M}
\end{equation}
and speed of sound $c_s$~\cite{gravdynam}
\begin{equation}
 c_s^2=\frac{g\rn+\lambda}{m}.\label{eq:cs}
\end{equation}

Finally, by using the mode normalization~\eqref{eq:normuv}, the system~\eqref{eq:uvequations} and the dispersion relation~\eqref{eq:disp}, it is possible to compute $\uk$ and $\vk$:
\begin{equation}
 \uk^2=\frac{1}{1-D_\bk^2},
 \qquad
 \vk^2=\frac{D_\bk^2}{1-D_\bk^2},
 \label{eq:bogocoeff}
\end{equation}
where both $\uk$ and $\vk$ are chosen to be real and
\begin{equation}
 D_\bk\equiv\frac{\hbar\omega-\left(\hbar^2\bk^2/2m+g\rn+\lambda\right)}{g\rn-\lambda}.\label{eq:dk}
\end{equation}
%

\subsection{Analogue gravitational dynamics}	%
\label{subsec:gravdyn}				%

When the homogeneous condensate background is perturbed by small inhomogeneities, the Hamiltonian for the quasi-particles can be written as (see~\cite{gravdynam})
\begin{equation}\label{eq:nonrelhquasip}
\hH_{\rm quasip.} \approx {\cal M} c_s^2- \frac{\hbar^2 \nx^2}{2{\cal M}} + {\cal M}\Phi_{\rm g}.
\end{equation}
$\hH_{\rm quasip.}$ is the non-relativistic Hamiltonian for particles of mass ${\cal M}$ [see Eq.~\eqref{eq:M}] in a gravitational potential
\begin{equation}\label{gravitationalpotential}
\Phi_{\rm g}(\bx) = \frac{(g\rn+3\lambda)(g\rn+\lambda)}{2\lambda m} \delta\rho(\bx)
\end{equation}
and $\delta\rho(\bx)=[(\rn(\bx)/\rho_\infty)-1]/2$, where $\rho_\infty$ is the asymptotic density of the condensate. 
Moreover, the dynamics of the potential $\Phi_{\rm g}$ is described by a Poisson-like equation
\begin{equation}
 \left[\nx^2-\frac{1}{L^2}\right]\Phi_{\rm g}=4\pi G_{N}\rho_{\rm p}+C_\Lambda,\label{eq:poisson}
\end{equation}
which is the equation for a non-relativistic short-range field with length scale
\begin{equation}\label{eq:L}
 L=\frac{a}{\sqrt{16\pi\rn a^3}}
\end{equation}
and gravitational constant
\begin{equation}
 G_{\rm N}=\frac{g(g\rn+3\lambda)(g\rn+\lambda)^2}{4\pi\hbar^2m\lambda^{3/2}(g\rn)^{1/2}}.
\end{equation}
Despite the obvious difference between $\Phi_{\rm g}$
and the usual Newtonian gravitational potential, we insist in calling it the {\it Newtonian potential}
because it enters the acoustic metric exactly as
the Newtonian potential enters the metric tensor  in the Newtonian limit of GR.

The source term in Eq.~\eqref{eq:poisson} contains both the contribution of real phonons (playing the role of matter)
\begin{equation}
 \rho_{\rm p}  = {\cal M}\rn\left[\left(\sev{\hpd\hp}-\vev{\hpd\hp}\right)+
 \frac{1}{2}\re\left(\sev{\hp\hp}-\vev{\hp\hp}\right)\right],\label{eq:rhophonons}
\end{equation}
where $|\zeta\rangle$ is some state of real phonons and $|0\rangle$ is the Fock vacuum of the quasiparticles ($\ak|0\rangle=0, \: \forall\,\bk$), as well as a cosmological constant like term (present even in the absence of phonons/matter)
\begin{equation}\label{eq:Clambda}
 C_\Lambda=\frac{2g\rn (g\rn+3\lambda)(g\rn+\lambda)}{\hbar^2\lambda}
 \re \left[\vev{\hpd\hp}+\frac{1}{2}\vev{\hp\hp}\right].
\end{equation}
Note that the source term in the correct weak field approximation of Einstein's equations is $4\pi G_{N}(\rho+3p/c^2)$. For standard nonrelativistic matter, $p/c^2$ is usually negligible with respect to $\rho$. However, it cannot be neglected for the cosmological constant, since $p_\Lambda/c^2=-\rho_\Lambda$. As a consequence the analogue cosmological constant is
\begin{equation}\label{eq:lambdacc}
 \Lambda=-\frac{C_\Lambda}{2c_{s}^{2}}.
\end{equation}

\subsection{BEC ground state energy}	%
\label{subsec:groundbec}		%

\begin{figure}
\centering
\includegraphics[width=0.8\textwidth]{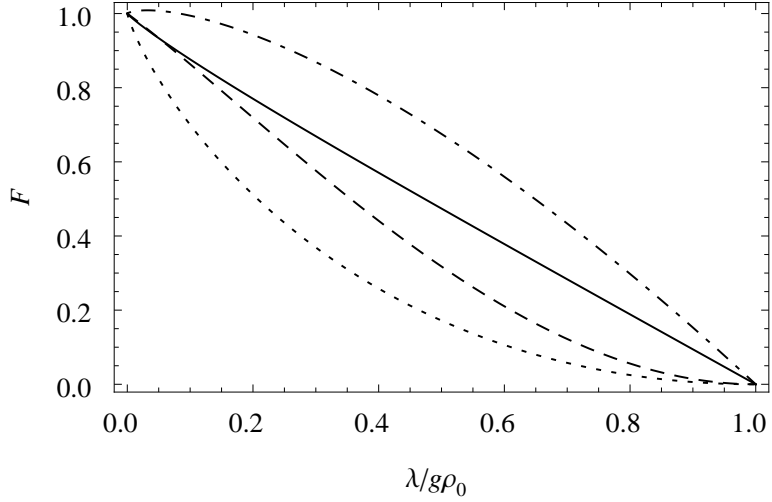}
\caption{\label{fig:fs}$F_h$ [dashed line, Eq.~\eqref{eq:h2vev2}], $F_\rho$ [dotted line, Eq.~\eqref{eq:depletion}], $F_{\phi\phi}$ [dot-dashed line, Eq.~\eqref{eq:phiphi}], and $F_{\Lambda}$ [solid line, Eq.~\eqref{eq:cosmconst}].}
\end{figure}

By using the formalism developed in Sec.~\ref{subsec:settings}, we compute the vacuum expectation value of $\cal\hH$ in the ground state $|0\rangle$. To this aim, it is convenient to use the expansion of $\cal\hH$ in powers of $\hp$ given in Eq.~\eqref{eq:Hexpansion}.
The energy density $h_0$ of the condensate (density of ${\cal H}_0$) in an homogeneous condensate is straightforwardly obtained from Eq.~\eqref{eq:h0U1breaking}, using the relation~\eqref{eq:mu} between the chemical potential $\mu$ and the couplings $g$ and $\lambda$
\begin{equation}
 h_0=-\frac{g\rn^2}{2}.\label{eq:h0vev}
\end{equation}
The expectation value of ${\cal\hH}_1$ vanishes because it contains only odd powers of the phonon field $\hp$. Finally, the density $h_2$ of the expectation value of ${\cal\hH}_2$ can be computed by using the equation of motion of the perturbations~\eqref{eq:hp}, such that the expression of ${\cal\hH}_2$ simplifies to
\begin{equation}
{\cal\hH}_2=\frac{\ii\hbar}{2}\int\! \dx\,\rn\left[\hpd\pdt\hp-(\pdt\hpd)\hp\right].
\end{equation}
Inserting the field expansions~\eqref{eq:phiexpansionk} in the above expression, one obtains
\begin{equation}
 {\cal\hH}_2=\int\!\dk\, \hbar\om\left[\akd\ak-\int\!\frac{\dx}{(2\pi)^2}|\vk|^2\right],
\end{equation}
which implies that the density of $\vev{{\cal\hH}_2}$ is
\begin{equation}
 h_2=-\int\!\frac{\dk}{(2\pi)^3}\hbar\omega|\vk|^2.\label{eq:h2vev}
\end{equation}
The integral in Eq.~\eqref{eq:h2vev} is computed by using Eqs.~\eqref{eq:bogocoeff} and~\eqref{eq:dk}.
The divergence in $d=3$ spatial dimensions is regularized by performing the calculation with $d<3$ and then going to the limit $d\to3$. This regularization is equivalent to the subtraction of higher order interaction terms (see also~\cite{huang})
\begin{equation}\label{eq:h2vev2}
 h_2=\frac{64}{15\sqrt{\pi}}g\rn^2\sqrt{\rn a^3}\,\,F_h\!\left(\frac{\lambda}{g\rn}\right),
\end{equation}
where $a=4\pi gm/\hbar^2$ is the scattering length, $F_h$ is plotted in Fig.~\ref{fig:fs} (dashed line) and $F_h(0)$=1.
The total grand canonical energy density is therefore
\begin{equation}\label{eq:grandpotential}
 h=h_0+h_2=\frac{g\rn^2}{2}\left[-1+\frac{128}{15\sqrt{\pi}}\sqrt{\rn a^3}\,\,F	_h\!\left(\frac{\lambda}{g\rn}\right)\right].
\end{equation}
%

To compute the energy density $\epsilon=h+\mu\rho$, that is the density of $\vev{\hH}=\vev{{\cal H}+\mu\hat N}$), we have to express at first the density of condensed atoms $\rn$ in terms of the total number density $\rho$.
To this aim, we expand the particle number operator $\hat N$ in powers of $\hp$ 
\begin{equation}
 \hat N=N_0+\hat N_1+\hat N_2,
\end{equation}
where, as in Eq.~\eqref{eq:Hexpansion}, $N_0$, $\hat N_1$, and $\hat N_2$ contain respectively no power of $\hp$, only first powers, and only second powers
\begin{align}
 N_0&=\int\!\dx\,\rn,\displaybreak[0]\\
 \hN_1&=\int\!\dx\,\Pn\,\hpd+\mbox{h.c.},\displaybreak[0]\\
 \hN_2&=\int\!\dx\,\rn\,\hpd\hp.\label{eq:N2}
\end{align}
The density of $N_0$ is then simply
\begin{equation}
 \rho_0=|\Pn|^2,
\end{equation}
$\vev{\hat N_1}$ vanishes, and the density of $\vev{\hat N_2}$ is
\begin{equation}\label{eq:depletion}
 \rho_2=\!\int\!\!\frac{\dk}{(2\pi)^3}|\vk|^2=\frac{8\rn}{3\sqrt{\pi}}\sqrt{\rn a^3}\,\,F_\rho\!\left(\frac{\lambda}{g\rn}\right),
\end{equation}
where $F_\rho$ satisfies $F_\rho(0)=1$ (see Fig.~\ref{fig:fs}, dotted line).
This is the number density of non-condensed atoms ({\it depletion}). Note that $\rn a^3$ is the so called dilution factor which has to be much smaller than 1 for the Hamiltonian~\eqref{eq:HU1breaking} to hold.

Furthermore, when $\lambda=0$, inverting the expression for total particle density, $\rho=\rn+\rho_2$, one obtains, up to the first order in $\sqrt{\rho a^3}$
\begin{equation}
 \rn=\rho\left[1-\frac{8}{3\sqrt{\pi}}\sqrt{\rho a^3}\right],
\end{equation}
which is the density of condensed atoms in terms of the total density~$\rho$ and the scattering length~$a$~\cite{lhy}. In this case, $\mu=g\rn$, such that the energy density is
\begin{equation}
 \epsilon=h+\mu\rho= \frac{g\rho^2}{2}\left[1+\frac{128}{15\sqrt{\pi}}\sqrt{\rho a^3}\right].\label{eq:energy}
\end{equation}
This is the well known Lee--Huang--Yang~\cite{lhy} formula for the ground state energy in a condensate at zero temperature.
In general, when the $U(1)$ breaking term is small, this term is expected to be the dominant contribution to the ground state energy of the condensate.

\subsection{What does the cosmological constant correspond to?}	%
\label{subsec:cosmobec}						%

We shall now compare the energy density and the grand canonical energy density found in the previous section with the effective cosmological constant $\Lambda$ of Eq.~\eqref{eq:lambdacc}.
$C_\Lambda$ of Eq.~\eqref{eq:Clambda} is computed by using Eq.~\eqref{eq:depletion} and the following expectation value
\begin{equation}\label{eq:phiphi}
 \vev{\hp\hp}=\!\int\!\!\frac{\dk}{\rn(2\pi)^3}\uk\vk=\frac{8}{\sqrt{\pi}}\sqrt{\rn a^3}\,F_{\phi\phi}\!\left(\frac{\lambda}{g\rn}\right),
\end{equation}
where $F_{\phi\phi}(0)=1$ (see Fig.~\ref{fig:fs}, dot-dashed line). We finally obtain
\begin{equation}\label{eq:cosmconst}
 \Lambda=-\frac{20m\,g\rn\,(g\rn+3\lambda)}{3\sqrt{\pi}\hbar^2\lambda}\sqrt{\rn a^3}\,F_\Lambda\!\left(\frac{\lambda}{g\rn}\right),
\end{equation}
where $F_\Lambda=(2F_\rho+3F_{\phi\phi})/5$ (see Fig.~\ref{fig:fs}, solid line).

Let us now compare the value of $\Lambda$ either with the ground-state grand canonical energy density $h$ [Eq.~\eqref{eq:grandpotential}], which was suggested in~\cite{volovik1,volovik2} as the correct vacuum energy corresponding to the cosmological constant, or to the ground-state energy density $\epsilon$ of Eq.~\eqref{eq:energy}.
Evidently, $\Lambda$ does not correspond to either of them: even when taking into account the correct behavior at small scales, the vacuum energy computed with the phonon EFT does not lead to the correct value of the cosmological constant appearing in Eq.~\eqref{eq:poisson}.
Noticeably, since $\Lambda$ is proportional to $\sqrt{\rn a^3}$, it can even be arbitrarily smaller both than $h$ and than $\epsilon$, if the condensate is very dilute. Furthermore, $\Lambda$ is proportional only to the subdominant second order correction of $h$ or $\epsilon$, which is strictly related to the depletion [see Eq.~\eqref{eq:depletion}].

Furthermore, several scales show up in the emergent system, in addition to the na\"ive Planck scale computed by combining the emergent constants $G_{\rm N}$, $c_s$ and $\hbar$:
\begin{equation}
 L_{\rm P}=\sqrt{\frac{\hbar c_s^5}{G_{\rm N}}}\propto \left(\frac{\lambda}{g\rn}\right)^{-3/4}(\rn a^3)^{-1/4} a.
\end{equation}
For instance, the Lorentz-violation scale (\ie, the healing length of the condensate)
\begin{equation}
 L_{\rm LV}=\xi \propto(\rn a^3)^{-1/2} a
\end{equation}
differs from $L_{\rm P}$, suggesting that the breaking of the Lorentz symmetry might be expected at a much longer scale than the Planck length (much smaller energy than the Planck energy), since the ratio $L_{\rm LV}/L_{\rm P}\propto(\rn a^3)^{-1/4}$ increases with the diluteness of the condensate.

To conclude, it is instructive to compare the energy density corresponding to $\Lambda$
\begin{equation}
{\cal E}_\Lambda=\frac{\Lambda c_s^4}{4\pi G_{\rm N}}
\end{equation}
to the na\"ive Planck energy density
\begin{equation}
 {\cal E}_{\rm P}=\frac{c_s^7}{\hbar G_{\rm N}^2}.
\end{equation}
The former is much smaller than the value computed from zero-point-energy calculations with a cut off at the Planck scale. Indeed, the ratio between these two quantities
\begin{equation}
 \frac{{\cal E}_\Lambda}{{\cal E}_{\rm P}}\propto \rn a^3\left(\frac{\lambda}{g\rn}\right)^{-5/2}
\end{equation}
is again controlled by the diluteness parameter $\rn a^3$.

\subsection{The spinor BEC case}
From the previous analysis one can easily realize that $L_{\rm LV}$ scales with $\rn a^3$ exactly as the range of the gravitational force [see Eq.~\eqref{eq:L}], signaling that this model is too simple to correctly grasp all the desired features. However,
in more complicated systems~\cite{nbec}, this pathology can be cured, in the presence of suitable symmetries, leading to long range potentials.

In the case of a BEC system containing several different bosonic species, the so called spinor BEC (see \cite{calzetta} in these proceedings for a discussion of their relevance as analogue models and for references) it is possible to repeat the analysis that we have described here, with minimal variations in the method. There are significant differences, however. In general, without tuning or symmetries (internal symmetries among the various components, essentially) the geometrical structure describing the propagation of phonons is not a Lorentzian metric, at low energy, but rather something like a Finsler structure \cite{silke2BEC}. Additionally, the coupling  between phonons and the condensate wavefunctions
becomes extremely nontrivial, with the analogue Newtonian potential obeying some complicated equation that cannot be cast in the form of a Poisson equation, even including a Yukawa mass term. See \cite{nbec} for the complete discussion.

However, if one imposes a symmetry in the system, that is, if there is an underlying symmetry under the permutation of the species, the situation improves so much that indeed a realistic analogue can be obtained. Consider a system with $N$ components (whose nature will be neglected, for the present reasoning), labelled by roman letters $A,B=1,...,N$. A Hamiltonian obeying the requirement just mentioned would be of the form:
\begin{equation}
\hat{H} = \int d^3x \sum_{A=1}^{N} \hat{\Psi}_{A}\left( -\frac{\hbar^2}{2m} - \mu -V(x) +\frac{g}{2} |\hat{\Psi}_A|^2 +\frac{g'}{2} \sum_{B\neq A}|\hat{\Psi}_B|^2 \right)\hat{\Psi}_A
\end{equation}
where $\mu, \lambda,g$ are playing the same role of the corresponding quantities defined in \eqref{eq:HU1breaking}.  
In this case, due to the richer structure available, one can recover a low energy notion of Lorentz invariance, and a distinguished long range analogue Newtonian potential. 

{In particular, one can describe the deviations from homogeneity in each component of the condensate by expanding the wave function as
\begin{equation}
\left[\Psi_0\right]_A=\sqrt{[\rho_0]_A}+\alpha_A(x)+{\rm i} \beta_A(x)\,,
\end{equation} 
where $\alpha(x),\beta(x)$ are real functions. Within this ansatz it is then possible to show (see again~\cite{nbec}) that a long range Newtonian potential of the form
\begin{equation}
\Phi_{N}(x) \propto \sum_{A=1}^N \beta_A(x).
\end{equation}
can exist and that this potential is indeed coupled to several (massive) phonons in a universal way, hence manifesting the emergence of an analogue of the equivalence principle.

Therefore, the fact that in the single BEC there seems to be a basic flaw given by the short range potential should be seen as an artifact of the simplicity of the model, and not a basic obstruction. The above mentioned  generalized model can remove this unpleasant feature, allowing the construction of a more realistic dynamical analogue.


Most importantly, the very same analysis that has been reported here can be done for the multi-component case so to  study the nature of the vacuum contribution to the source term of the analogue Poisson equation. Even without reporting a full calculation one can easily see that the bulk of the result will be of the same nature: the cosmological term will be related to the depletion factor, and hence non necessarily Planckian (in the sense discussed above).
}

\section{Summary and Conclusions}	%
\label{sec:conclusionscosmobec}		%

Let us summarize the results.
We have investigated the problem of calculating the cosmological constant through an analogy with condensed matter. As a model we adopted the BEC with $U(1)$-symmetry breaking proposed in~\cite{gravdynam}, given that in this system it is possible to define the analogue of the gravitational field satisfying a modified Poisson-like equation. The source term appearing in this equation is made of two pieces: the first one is related to real phonons, corresponding to matter fields, in the analogy, and that generates the gravitational field associated to the presence of clumps of matter. The second one represents a vacuum contribution, that precisely matches a cosmological constant in the Newtonian regime.

Of course this is just a toy model for gravity, which nevertheless provides a clear description of the procedure that should be followed to correctly compute the cosmological constant. We showed that the analogue cosmological constant cannot be computed as the total zero-point energy of the condensed matter system, even when taking into account the natural cut-off coming from the knowledge of the correct microphysics, as suggested in~\cite{volovik1,volovik2}. In fact, the value of $\Lambda$ is related only to a part of the zero-point energy, namely a subleading term proportional to the quantum depletion of the condensate.%
\footnote{In a comment~\cite{volovikcomm} to~\cite{cosmobec}, an alternative point of view has been proposed. There it has been argued that the difference between the zero-point energy and the actual value of the cosmological constant is smaller, the closer is the analogue system to reproduce GR.}

Let us further elaborate on the significance of this result. The depletion factor is intimately related to the fact that the Fock vacuum of the fundamental bosons and the Fock vacuum of the phonons are inequivalent, a fact that is more profound that just an energy consideration. It is a statement about
the full quantum state corresponding to the condensed phase, which is the regime in which we can speak about a semiclassical gravity analogue. It is a quantity that encodes in a specific way the information that the system is in a BEC phase, and not just a generic many body state.

This observation, taken alone, would be of little significance outside a condensed matter community. However, it assumes a different relevance if we move to the perspective of a quantum gravity model. There, part of the properties of
the long range/low energy/continuum/semiclassical regime that will result in a gravitational theory
will be due to the microphysics, on one hand, and to the particular regime or considered state that will encode the fact that the model is considered in the long range/low energy/continuum/semiclassical limit. The knowledge of the microscopic dynamics must then be supplemented by the specific information about the considered state.
In this perspective, a similar situation arises within loop quantum gravity models \cite{Alexander1,Alexander2}, suggesting a BCS energy gap as an origin for the cosmological constant. When we say that the cosmological constant has to be computed in terms of the microscopic theory, we also imply that we have identified the state or the class of states that will correspond to a semiclassical state, and that the effective dynamics will be the outcome of the microscopic dynamics as well as of the state considered.

Going back to our model, this result suggests a twofold interpretation. First, there is no {\it a priori}\/ reason why the cosmological constant should be computed as the zero-point energy of the system, even when this energy is calculated correctly taking into account the corrections coming from the microphysics of the system. The computation of this constant must pass instead through the derivation of Einstein's equations emerging from the underlying microscopic system.
Second, the value of $\Lambda$ can be several orders of magnitude smaller than the total vacuum energy density, depending on the features of the fundamental structure from which the spacetime emerges. The fine-tuning problem would therefore become much less worrying. However, no indication about the coincidence problem comes from the analysis of this analogy.

Moreover, even in such a simple system, several different scales show up.
For instance, the Planck scale computed with the fundamental dimensionful constants is very different from the scale at which Lorentz symmetry is broken ({\it i.e.}\/ when the dispersion relation is no longer linear), which in this case coincides with the range of gravitational interaction. This is just a coincidence due to the extreme simplicity of the model, and we have argued that a spinor BEC might alleviate this problem. In general, these scales are functions of the fundamental scattering length and of the diluteness of the condensate. Most importantly, the energy density associated with the cosmological constant can be much smaller than the value calculated with a cut off at the Planck scale, being proportional to the diluteness parameter of the condensate.
Of course, this result must not be strictly translated to the gravity side of the analogy. It shows nevertheless how the problem of defining scales for possible Lorentz violations in relation to the so called Planck units is very far from being trivial. Furthermore, a very wide range of options is left open, depending on the fundamental structure the spacetime is emerging out. In particular, since the value of the cosmological constant appears to be strongly dependent on the granular structure of the spacetime, the comparison with the observed value of this constant might represent an important test for the validity of any theory of quantum gravity.

Actually, we can further broaden the scope of the discussion.
Our result strongly supports a picture where gravity is a collective phenomenon in a pregeometric theory (for a review of different ideas and references, see \cite{emergent}).
In fact, the cosmological constant puzzle is more easily addressed in those scenarios.
From an emergent gravity approach, the low energy effective action (and its renormalization group flow) for gravity and matter fields is obviously computed within a framework that has nothing to do with QFT in curved spacetime.

In these scenarios, if we interpreted the cosmological constant as a coupling constant controlling some self-interaction of the gravitational field (\ie~putting it on the LHS of Einstein equations), rather than as a vacuum energy (on the RHS), it would straightforwardly follow that the explanation of its value (and of its properties under renormalization) would naturally sit outside the domain of semiclassical gravity. From this point of view, the cosmological constant works as a phenomenological parameter controlling the constitutive relation that controls the self-interaction of the gravitational field as it is induced from (and at the same time it summarizes) the underlying dynamics. As such, its value cannot be computed within the semiclassical gravity regime, but has to be computed on the basis of the underlying microscopic dynamics.

In this respect, it is conceivable that the very notion of cosmological constant as a form of energy intrinsic to the vacuum is ultimately misleading. This is the point of view of many quantum gravity models. An interesting case is represented by group field theories, a generalization to higher dimensions of matrix models for two dimensional QG \cite{oriti}.
To date, little is known about the macroscopic regime of them, conjectured to be some form of gravitational theory, even though some preliminary steps have been recently done~\cite{oritisindoni}.
In such a framework, it is transparent that the origin of the gravitational coupling constants has nothing to do with ideas like ``vacuum energy'' or statements like ``energy gravitates'', because energy {\it itself} is an emergent concept. Rather, the value of $\Lambda$ is determined by the microphysics, and, most importantly, by the procedure to approach the continuum semiclassical limit. For instance, in \cite{LoGFT} it is shown that the effective dynamics will be controlled essentially by the critical behavior of the model, as expected, up to the point in which the gravitational coupling constants for an effective field theory will be functions of the
critical exponents associated to the phase transition defining the continuum limit, \ie~to the state of the system that defines the continuum limit. Clearly these are concepts totally disconnected from notions like energy, fields, metric, etc. Most importantly, these critical exponents cannot be computed with field theoretic arguments in semiclassical gravity. The similarity to the case we are investigating here is manifest.

Lacking a full fledged derivation of semiclassical continuum gravity from a quantum gravity model,
toy models are playing a crucial role. While a detailed calculation cannot be avoided forever, analogue models can and will represent key assets, as sources of inspiration and techniques to address basic problems in quantum gravity. The simple model discussed here, involving a BEC in which an effective analogue dynamics for a gravitational field can be defined, shows how powerful this idea can be in elucidating the nature of certain puzzles, as well as concretely suggesting a possible way to approach the solution
in quantum gravity settings. 
In this respect, the reasoning of this chapter sheds a totally different light on the cosmological constant problem, turning it from a failure of EFT to a possible window on the process with which spacetime arises as an effective description.


\begin{thebibliography}{10}



\bibitem{carroll}
S.~M. Carroll,
\textit{The cosmological constant},
Living Rev. Relativity \textbf{4}, 1 (2001).

\bibitem{firstcosm}
A.~{Einstein}.
\textit{Kosmologische Betrachtungen zur allgemeinen Relativit{\"a}tstheorie
  (Cosmological Considerations in the General Theory of Relativity)},
Sitzungsbe. Preuss. Akad. Wiss. 142 (1917).


\bibitem{finetuning}
E.~{Cremmer}, S.~{Ferrara}, C.~{Kounnas}, and D.~V. {Nanopoulos},
\textit{Naturally vanishing cosmological constant in N=1 supergravity},
Phys. Lett. B \textbf{133}, 61 (1983).

\bibitem{rovelli}
E.~{Bianchi} and C.~{Rovelli},
\textit{Why all these prejudices against a constant?},
arXiv:1002.3966v3 [astro-ph.CO].


\bibitem{susy}
S.~P. {Martin},
\textit{A Supersymmetry Primer},
In \textit{Perspectives on Supersymmetry}, edited by {G.~L.~Kane}, 1 (1998).

\bibitem{Paddy}
T.~Padmanabhan,
\textit{Vacuum fluctuations of energy density can lead to the observed cosmological constant},
Class. Quant. Grav. \textbf{22} L107 (2005).

\bibitem{livrev}
C.~{Barcel{\'o}}, S.~{Liberati}, and M.~{Visser},
\textit{Analogue gravity},
Living Rev. Relativity \textbf{14}, 3 (2011).

\bibitem{volovik1}
G.~E. Volovik,
\textit{Vacuum Energy and Cosmological Constant: View from Condensed Matter},
J. Low Temp. Phys. \textbf{124}, 25 (2001).

\bibitem{volovik2}
G.~E. {Volovik},
\textit{Cosmological constant and vacuum energy},
Ann. Phys. \textbf{14}, 165 (2005).

\bibitem{volovikbook}
G.~E. Volovik,
\textit{The Universe in a Helium Droplet},
(Oxford University Press, Oxford, U.K., 2003).

\bibitem{gravdynam}
F.~{Girelli}, S.~{Liberati}, and L.~{Sindoni},
\textit{Gravitational dynamics in Bose--Einstein condensates},
Phys. Rev. D \textbf{78}, 084013 (2008).

\bibitem{cosmobec}
S.~{Finazzi}, S.~{Liberati}, and L.~{Sindoni},
\textit{The cosmological constant: a lesson from Bose--Einstein condensates},
Phys. Rev. Lett. \textbf{108} 071101 (2012).

\bibitem{weinbergcc}
S.~{Weinberg},
\textit{The cosmological constant problem},
Rev. Mod. Phys. \textbf{61}, 23 (1989).

\bibitem{birreldavies}
N.~D. {Birrell} and P.~C.~W. {Davies},
\textit{Quantum Fields in Curved Space},
(Cambridge University Press, Cambridge, U.K., 1984).

\bibitem{huang}
K.~{Huang},
\textit{Statistical Mechanics}.
(John Wiley \& Sons, New York, 1987), 2nd Edition.

\bibitem{nbec}
L.~{Sindoni},
\textit{Emergent gravitational dynamics from multi-Bose--Einstein-condensate hydrodynamics?},
Phys. Rev. D \textbf{83}, 024022 (2011).

\bibitem{ulf}
U.~{Leonhardt}, T.~{Kiss}, and P.~{{\"O}hberg},
\textit{Theory of elementary excitations in unstable Bose--Einstein condensates and the instability of sonic horizons},
Phys. Rev. A \textbf{67} 033602 (2003).

\bibitem{lhy}
T.~D. {Lee}, K.~{Huang}, and C.~N. {Yang},
\textit{Eigenvalues and Eigenfunctions of a Bose System of Hard Spheres and Its Low-Temperature Properties},
Phys. Rev. \textbf{106}, 1135 (1957).

\bibitem{calzetta}
 E.~Calzetta,
  ``Analog cosmology with spinor BECs,''
  arXiv:1111.5261 [gr-qc].

\bibitem{silke2BEC}
 S.~Liberati, M.~Visser and S.~Weinfurtner,
\textit{Analogue quantum gravity phenomenology from a two-component Bose-Einstein condensate},
  Class.\ Quant.\ Grav.\ \ {\bf 23} (2006) 3129
  [gr-qc/0510125].

\bibitem{volovikcomm}
G. Jannes, G.~E. Volovik,
\textit{The cosmological constant: a lesson from the Weyl superfluid $^3$He-A},
arXiv:1108.5086 [gr-qc].

\bibitem{Alexander1}
S.~H.~S. {Alexander} and G.~{Calcagni},
\textit{Quantum Gravity as a Fermi Liquid},
Found. Phys. \textbf{38}, 1148 (2008).

\bibitem{Alexander2}
S.~H.~S. {Alexander} and G.~{Calcagni},
\textit{Superconducting loop quantum gravity and the cosmological constant},
Phys. Lett. B \textbf{672}, 386 (2009).

\bibitem{emergent}
 L.~Sindoni,
\textit{Emergent models for gravity: an overview of microscopic approaches}, 
SIGMA \textbf{8}, 027 (2012) . 
  
\bibitem{oriti}
 D.~Oriti,
 \textit{The microscopic dynamics of quantum space as a group field theory},
  arXiv:1110.5606 [hep-th].

\bibitem{oritisindoni}
D.~{Oriti} and L.~{Sindoni},
\textit{Toward classical geometrodynamics from the group field theory hydrodynamics},
New J. Phys. \textbf{13}, 025006 (2011).

\bibitem{LoGFT}
  L.~Sindoni,
\textit{Gravity as an emergent phenomenon: a GFT perspective},
  arXiv:1105.5687 [gr-qc].

\end{thebibliography}

\end{document}